\documentclass[12pt]{article}
\usepackage{amssymb,epsfig}

\renewcommand{\thefootnote}{\fnsymbol{footnote}}

\begin{document}

\title{
\begin{flushright}
\ \\*[-80pt] 
\begin{minipage}{0.2\linewidth}
\normalsize
YITP-08-83 \\
TU-828 \\
KUNS-2161 \\*[50pt]
\end{minipage}
\end{flushright}
{\Large \bf 
Duality cascade of softly broken 
supersymmetric theories
\\*[20pt]}}

\author{
\hspace*{-1.5cm}
Hiroyuki~Abe$^{1}$
, \ 
Tetsutaro~Higaki$^{2}$
, \ 
Tatsuo~Kobayashi$^{3}$
, \\
\hspace*{-1.5cm}
Kazutoshi~Ohta$^{2}$
, \ 
Yuji~Omura$^{3}$
\ and \ 
Haruhiko~Terao$^{4}$
\\*[20pt]
\hspace*{-1cm}
$^1${\it \normalsize 
Yukawa Institute for Theoretical Physics, Kyoto University, 
Kyoto 606-8502, Japan} \\
\hspace*{-1cm}
$^2${\it \normalsize 
Department of Physics, Tohoku University, 
Sendai 980-8578, Japan} \\
\hspace*{-1cm}
$^3${\it \normalsize 
Department of Physics, Kyoto University, 
Kyoto 606-8502, Japan} \\
\hspace*{-1cm}
$^4${\it \normalsize 
Department of Physics, Nara Women's University, 
Nara 630-8506, Japan} \\*[50pt]}

\date{
\centerline{\small \bf Abstract}
\begin{minipage}{0.9\linewidth}
\medskip 
\medskip 
\small
We study the duality cascade of softly broken supersymmetric theories.
We investigate the renormalization group (RG) flow of 
SUSY breaking terms as well as supersymmetric couplings.
It is found that the magnitudes of SUSY breaking terms are 
suppressed in most regimes of the RG flow through 
the duality cascade.
At one stage of cascading, 
the gaugino mass of the strongly coupled sector and 
scalar masses converge to certain values, which are determined by 
the gauge coupling and the gaugino mass of the weakly coupled sector.
At the next stage, the strongly and weakly coupled sectors are 
interchanged with each other.
We also show the possibility that cascading would be terminated 
by the gauge symmetry breaking, which is induced by 
the so-called B-term.
\end{minipage}
}

\begin{titlepage}
\maketitle
\thispagestyle{empty}
\end{titlepage}


\renewcommand{\thefootnote}{\arabic{footnote}}
\setcounter{footnote}{0}

\section{Introduction}

Conformal dynamics plays important roles in 
various aspects of (supersymmetric) field theories 
and particle phenomenology.
Conformal fixed points and conformal field theories (CFTs)
are essential in Seiberg duality \cite{Seiberg:1994pq,Intriligator:1995au}.
That leads to more complicated and 
interesting renormalization group (RG) flows of 
dual field theories, that is, 
the duality cascade \cite{Klebanov:2000hb,Strassler:2005qs},
which is a successive chain of the dualities from 
the ultraviolate (UV) region to the infrared (IR) region and
reduces the rank of gauge groups one after another.
Furthermore, the AdS/CFT (gravity/gauge) correspondence
\cite{Maldacena:1997re}
suggests that the cascading theories would be realized
in supergravity theory with a warped background, that is, 
the Klebanov-Strassler warp throat.
In the supergravity description, the energy scale of the field theory
corresponds to the distance from a tip of the throat.
The duality cascade process means that the charges of D-branes
disappear as the prove brane gets closer to the tip.
The investigation of the duality cascade from the string/supergravity 
viewpoint is highly non-trivial check for the gravity/gauge correspondence.

Superconformal dynamics is also important in 
applications to particle phenomenology.
For example, conformal dynamics may generate 
realistic hierarchies of quark and lepton masses 
\cite{Nelson:2000sn,Kobayashi:2001kz,Kobayashi:2001is}.
Conformal dynamics has significant effects on 
supersymmetry (SUSY) breaking terms, too.
In simple gauge theories, soft SUSY breaking terms, 
i.e. the gaugino mass and soft scalar masses, are 
exponentially suppressed toward the IR  
attractive fixed point \cite{Karch:1998qa,Kobayashi:2001kz,Terao:2001jw}.
Thus, sfermion masses are exponentially suppressed 
in the above models with CFT-induced Yukawa hierarchy \cite{Nelson:2000sn,Kobayashi:2001kz,Kobayashi:2001is}.\footnote{
Five-dimensional warped theory with the same behavior was 
studied e.g. in \cite{Choi:2003di}.}
Another aspect of conformal dynamics relevant to SUSY breaking terms is 
conformal sequestering 
\cite{Luty:2001jh,Dine:2004dv,Sundrum:2004un,Ibe:2005pj,Schmaltz:2006qs,
Murayama:2007ge}.
Conformal dynamics may suppress flavor-dependent SUSY breaking terms 
and make flavor-blind contributions dominant.
Conformal dynamics may be important to realize 
a SUSY breaking model \cite{Abe:2007ki}.

Thus, superconformal dynamics is important in particle 
physics.
Here we study more about the duality cascade.
Recently, several models have been proposed to realize 
supersymmetric standard models as well as 
their extensions at the bottom of the cascade 
\cite{Cascales:2005rj,Heckman:2007zp}.
If we would like to realize the gauge theories in Type IIA/IIB string theory,
it admits high ranks of the gauge groups since
there are configurations with the various number of the D-branes.
To explain how we obtain the standard model like theories 
with fewer ranks from the infinitely many string vacua with large ranks,
those models are quite interesting and have opened 
possible candidates for high energy theories.
Those models are exactly supersymmetric.
At any rate, supersymmetry is broken in Nature 
even if supersymmetric theory is realized at high energy.
Thus, if the cascading theories are relevant to the particle physics 
at the weak scale, 
supersymmetry should be broken at a certain stage, e.g. 
at the top or bottom of the cascade (high or low energy) or 
between them (intermediate energy).
Here we assume that SUSY is softly broken at the beginning of 
the cascade.
Then, we study RG flows of SUSY breaking terms as well as 
supersymmetric couplings.

This paper is organized as follows.
In section \ref{sec:rigid-SUSY}, we review briefly the RG flow of 
supersymmetric couplings in the duality cascade.
In section \ref{sec:SUSY-breaking}, we study RG flows of SUSY breaking terms.
In section \ref{sec:symm-br}, we study symmetry breaking due to the B-term
by using illustrative examples. 
Section \ref{sec:conclusion} is devoted to conclusion and discussion.

\section{RG flow in duality cascade of 
rigid supersymmetric theories}
\label{sec:rigid-SUSY}

Here, we give a brief review on 
the RG flow in the duality cascade of rigid supersymmetric 
theories \cite{Klebanov:2000hb,Strassler:2005qs}.
We consider the gauge group $SU(kN) \times SU((k-1)N)$ 
and we denote their gauge couplings, $g_k$ and $g_{k-1}$.
Also, our model has two chiral multiplets $Q_r$ $(r=1,2)$ 
in the bifundamental representation of $SU(kN) \times SU((k-1)N)$, 
i.e. the fundamental representation for $SU(kN)$ and 
the anti-fundamental representation for $SU((k-1)N)$, 
and two chiral multiplets $\bar Q_s$ $(s=1,2)$ 
in the anti-bifundamental representation.
Then we introduce the following superpotential,
\begin{equation}
\label{eq:W-4}
W=h~{\rm tr} \det_{r,s} (Q_r \bar Q_s)=h\left[ 
(Q_1)^\alpha_a(\bar Q_1)^a_\beta (Q_2)^\beta_b (\bar Q_2)^b_\alpha - 
(Q_1)^\alpha_a(\bar Q_2)^a_\beta (Q_2)^\beta_b (\bar Q_1)^b_\alpha
\right],
\end{equation}
where the indices $\alpha$ and $\beta$ are group indices for 
$SU(kN)$ and the indices $a$ and $b$ are group indices for 
$SU((k-1)N)$.

Now, we study the RG flow of gauge couplings $g_k$ and 
$g_{k-1}$ and the quartic coupling $h$ and their 
fixed points.
The fields $Q_r$ and $\bar Q_s$ have the same anomalous dimension, 
which we denote by $\gamma_Q$.
In the NSVZ scheme \cite{Novikov:1983uc}, 
beta-function of the gauge coupling $g$ in generic gauge theory 
is  written as 
\begin{equation}
\mu \frac{d \alpha}{d\mu}=\beta_\alpha = -F(\alpha)[3T_G - \sum_i T_i(1-\gamma_i)],
\end{equation} 
where $\alpha=g^2/(8\pi^2)$ and 
\begin{equation}
F(\alpha) =  \frac{\alpha^2}{1-T_G \alpha}.
\end{equation} 
Here,  $T_i$ and $\gamma_i$ denote 
Dynkin indices and anomalous dimensions 
of the chiral matter fields, 
while $T_G$ denotes the Dynkin index of 
the adjoint representation.
For example, we have $T_G=N$ for the $SU(N)$ gauge group and 
$T_i=1/2$ for the fundamental representation of the $SU(N)$ gauge group.
Using this scheme, beta-functions of the gauge couplings $g_k$ and 
$g_{k-1}$ are written as  
\begin{eqnarray}
\beta_{\alpha_k} &=& -F(\alpha_k)N[k+2+2(k-1)\gamma_Q], \\
\beta_{\alpha_{k-1}} &=& -F(\alpha_{k-1})N[k-3+2k\gamma_Q].
\end{eqnarray}
In addition, we can write the beta-function of $\eta = h \mu$ as 
\begin{equation}
\beta_\eta = \eta (1+2\gamma_Q).
\end{equation}

Suppose that both $SU(kN)$ and $SU((k-1)N)$ sectors are within the conformal 
window \cite{Seiberg:1994pq}, 
i.e. $3k/2 \leq 2(k-1) \leq 3k$ and $3(k-1)/2 \leq 2k \leq 3(k-1)$.
Then, we have two fixed points 
\cite{Banks:1981nn,Seiberg:1994pq,Intriligator:1995au}, 
\begin{equation}
\label{eq:fp-k-1}
{\rm A}: ~~~ k-3+2k\gamma_Q = 0, \qquad \alpha_k=\eta=0,
\end{equation}
and 
\begin{equation}
\label{eq:fp-k}
{\rm B}: ~~~ k+2+2(k-1)\gamma_Q = 0, \qquad \alpha_{k-1}=\eta=0.
\end{equation}
The anomalous dimension $\gamma_Q$ is a function of the couplings.
We represent a value of the gauge coupling $g_{k-1}$ 
($g_k$) at the first (second) fixed point by 
$g_{k-1}^*$ ($g_{k}^*$).

At the vicinity of the first fixed point A given by
(\ref{eq:fp-k-1}) with $g_{k-1} \approx g_{k-1}^*$ and 
$0 < \alpha_k, \eta \ll 1$ (region I), 
it is found that $\beta_{\alpha_{k-1}} \approx 0$, 
$\beta_{\alpha_k} <0$ and $\beta_\eta >0$, that is, 
$\alpha_k$ increases and $\eta$ decreases toward the IR direction.
Thus, the theory would flow to the other fixed point 
B given by (\ref{eq:fp-k}) toward the IR direction.
On the other hand, 
around the fixed point B with 
$g_{k} \approx g_{k}^*$ and $0 < \alpha_{k -1}, \eta \ll 1$
(region II), 
it is found that $\beta_{\alpha_{k}} \approx 0$, 
$\beta_{\alpha_{k -1}} >0$ and $\beta_\eta <0$.
Hence, the quartic operator 
$h~{\rm tr} \det_{r,s} (Q_r \bar Q_s)$ is relevant and 
the coupling $\eta$ increases toward the IR, while 
$\alpha_{k-1}$ shrinks.

We could examine the RG flows of the gauge couplings
$\alpha_k$ and $\alpha_{k-1}$, if we admit using
the anomalous dimension $\gamma_Q$ obtained in the
1-loop level. 
For a sufficiently large $N$, the anomalous dimension $\gamma_Q$
is given as
\begin{equation}
\gamma_Q = -N(k\alpha_k + (k-1)\alpha_{k-1}) .
\end{equation}
In Fig.~1, the RG flows in the coupling space
$(\alpha_k, \alpha_{k-1})$ obtained in the NSVZ
scheme are shown in the case of $k=5$. 
Here we rescale the couplings as $N\alpha \to \alpha$.
The points A $(0, 0.05)$ and B $(0.175, 0)$ represent
the fixed points. The renormalized trajectory (R.T.) 
connecting these fixed points is shown by the bold line.

The flows in the region I are subject to
the conformal dynamics around the UV fixed point A, while
the flows in the region II are subject to that around the
IR fixed point B.
The convergence in the region I is not strong, since the
fixed point coupling $\alpha_{k-1}^*$ is not so strong
in the case of $k=5$.
It is seen in Fig.~1 that the R.T. bends on the way and 
the behavior of the R.T. changes quickly there.
Thus the RG property on the R.T. in Fig.~1 may be
characterized well as that in the region I or II. 

\begin{figure}[h]
\begin{center}
\includegraphics[width=0.65\textwidth]{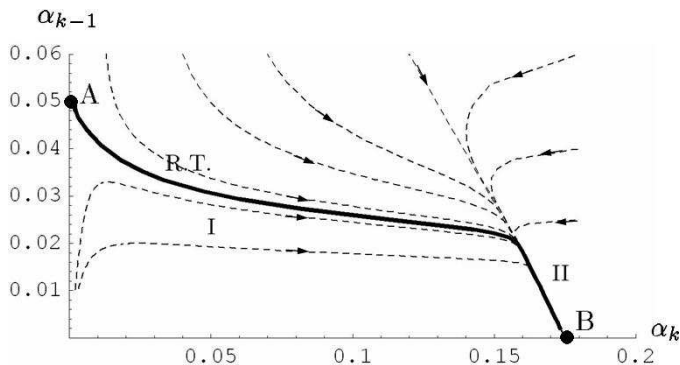}
\caption{RG flows in the coupling space 
$(\alpha_k, \alpha_{k-1})$ in the case of $k=5$. 
The points A and B represents
the UV and IR fixed points respectively. 
The renormalized trajectory connecting
these fixed points is shown by the bold line.
}
\end{center}
\end{figure}

The theory around the fixed point B 
is strongly coupled and would be well-described by 
its Seiberg dual \cite{Seiberg:1994pq,Intriligator:1995au}, 
which has the gauge group 
$SU((k-1)N) \times SU((k-2)N)$ and two bifundamental 
chiral multiplets $q_r$ and two anti-bifundamental chiral 
multiplets $\bar q_s$ and another kind of 
chiral multiplets $M_{rs}$, 
which correspond to $Q_r \bar Q_s$ and belong to  
the adjoint representation for $SU((k-1)N)$
\footnote{In the followings, we will ignore the irrelevant mesons 
$M^0_{rs}$ which are singlet for $SU((k-1)N)$.}
and 
singlet for $SU((k-2)N)$.
This dual theory has the following superpotential,
\begin{equation}\label{eq:W-dual}
W=y~{\rm tr}~\bar q_r M_{rs} q_s + m~{\rm tr} \det_{r,s} M_{rs}.
\end{equation}
The second term is the mass term of $M_{rs}$, which corresponds to 
$h~{\rm tr}~\det_{r,s}(Q_r \bar Q_s)$.
The mass $m$ would be related with the coupling $h$ as 
\begin{equation}
h(\Lambda_k) \Lambda_k \sim \frac{m(\Lambda_k)}{\Lambda_k},
\end{equation}
where $\Lambda_k$ is a typical energy scale of 
$SU(kN)$ gauge theory such as the energy scale, where 
the theory enters the conformal regime, i.e. 
$g_k(\Lambda_k) \approx g^*_k$.
The $\beta$-function of $\alpha_{k-2}$ is written 
in a way similar to those of $\alpha_{k-1}$ and $\alpha_{k}$.
In addition, the $\beta$-function of $y$ is written as 
\begin{equation}
\beta_y = \frac{y}{2}(\gamma_M + 2\gamma_q),
\end{equation}
where $\gamma_M$ is the anomalous dimension of $M_{rs}$.
The dual theory has a non-trivial fixed point, $g_{k-2}=g^*_{k-2}$ 
and $y=y^*$ when $g_{k-1}=0$, where $g^*_{k-2}, y^* \neq 0$.
At the fixed point, it is satisfied that $\gamma_M = - 2\gamma_q$, 
that is $M_{rs}$ has the same conformal dimension as 
$Q_r \bar Q_s$.
Thus, at the vicinity of the fixed point, $g_{k-2}=g^*_{k-2}$, 
$y=y^*$ and $g_{k-1}=0$, both operators, 
$h~{\rm tr}~\det_{r,s}(Q_r \bar Q_s)$ and 
$m~{\rm tr} \det_{r,s} M_{rs}$ are relevant, 
and the mass $m/\mu$ and coupling $h\mu$ increase towards the 
IR direction.
Because the fields $M_{rs}$ become heavy, we integrate out them 
and the effective superpotential results 
in \cite{Strassler:2005qs}\footnote{See also \cite{Kobayashi:2000di}.}
\begin{equation}
W = \tilde h~{\rm tr} \det_{rs} q_r \bar q_s,
\end{equation}
where $\tilde h = -y^2/m$.
The operator $\tilde h~{\rm tr} \det_{rs} q_r \bar q_s$ 
is irrelevant and the coupling $\tilde h$ decreases 
towards the IR direction.
Thus, the low energy effective theory is the same as the 
starting theory except replacing the gauge group 
$SU(kN) \times SU((k-1)N)$ by $SU((k-1)N) \times SU((k-2)N)$.
This RG flow would continue and the low-energy effective theory 
would become  the $SU((k-n)N) \times SU((k-n-1)N)$ gauge theory 
with the quartic superpotential 
$W = \tilde h~{\rm tr} \det_{rs} q_r \bar q_s$ until 
the theory becomes outside of the conformal window.
The RG flow toward the IR is illustrated as 
\begin{eqnarray}
\begin{array}{ccc}
(g_k\approx 0, g_{k-1} \approx g_{k-1}^*,\eta \approx 0) &
&   \\
\downarrow & & \\
(g_k\approx g_k^*, g_{k-1} \approx 0,\eta \approx 0) & 
\leftrightarrow & 
(g_{k -2}\approx g_{k -2}^*, g_{k-1} \approx 0,y \approx y^*, m/\mu
\approx 0) \\ \downarrow & {\rm dual} & \downarrow \\
(g_k\approx g_k^*, g_{k-1} \approx 0,\eta \gg 1) & \leftrightarrow & 
(g_{k -2}\approx g_{k -2}^*, g_{k-1} \approx 0,y \approx y^*, 
m/\mu \gg 1 ) \\
& & ~~~~~~~~~~~~~~~~~~~~~~~~~~ \downarrow {\rm integrating~out }~M_{rs} \\ 
& & (g_{k -2}\approx g_{k -2}^*, g_{k-1} \approx 0, \tilde \eta
\approx 0).
\\
\end{array} \nonumber
\end{eqnarray}

At the end of cascade we would obtain the 
$SU(2N) \times SU(N)$ gauge theory. 
The $SU(2N)$ gauge sector has the $2N$ flavors and 
the quantum deformed moduli space \cite{Seiberg:1994bz,Intriligator:1995au},
$\Delta W= X(\det_{\rm all}\, M_{rs} -B\bar B -\Lambda^{4N})$, 
where $X$ is a Lagrange multiplier superfield,
$B$ and $\bar B$ are baryon and anti-baryon superfields,
which are singlets under $SU(N)$. 
If we assume that only $B$ and $\bar B$ develop their vacuum expectation 
values (VEVs), 
then baryons and mesons become massive.
Thus the effective theory becomes the pure $SU(N)$ 
supersymmetric Yang-Mills theory and finally the theory is confined.

\section{RG flow of soft SUSY breaking terms}
\label{sec:SUSY-breaking}

Here, we study the RG flow of SUSY breaking terms 
in softly broken supersymmetric theories.
It is convenient to use the spurion method 
\cite{Yamada:1994id,Hisano:1997ua,Jack:1997pa,Kobayashi:1998jq,
ArkaniHamed:1998kj,Terao:2001jw} to derive 
RG equations of soft SUSY breaking terms from 
those for supersymmetric couplings.

\subsection{A simple theory with conformal fixed point}
\label{sec:3-1}

Here, we give a brief review on the spurion method and 
apply for a simple gauge theory with a conformal fixed point.
We consider a generic gauge theory with the gauge coupling $g$, 
the gaugino mass $M_{1/2}$, Yukawa couplings $y_{ijk}$, corresponding 
A-terms $a_{ijk}$ and soft scalar masses $m_i$.
We define the following superfield couplings 
\begin{eqnarray}
\tilde \alpha &=& \alpha \left( 1 + M_{1/2}\theta^2 + \bar M_{1/2} \bar \theta^2 
+ (2 |M_{1/2}|^2 + \Delta_g) \theta^2 \bar \theta^2 \right), \\
\tilde y_{ijk} &=& y_{ijk} - a_{ijk} \theta^2 + 
\frac{1}{2}(m_i^2+m_j^2+m_k^2)y_{ijk}\theta^2 \bar \theta^2,
\end{eqnarray}
where $\Delta_g$ is written as \cite{Kobayashi:1998jq}
\begin{equation}
\Delta_g = - \frac{F(\alpha)}{\alpha} 
\left[\sum_iT_i m_i^2 - T_G|M_{1/2}|^2
\right].
\end{equation}
Then, beta-functions of superfields $\tilde \alpha$ and
$\tilde y_{ijk}$ ($\bar{\tilde y}_{ijk}$) including soft SUSY breaking terms are 
obtained from those of $\alpha$ and $y_{ijk}$ (${\tilde y}_{ijk}$),
$\beta_\alpha(\alpha,y_{ijk},\bar{y}_{ijk})$ and $\beta_{y_{ijk}}(\alpha,y_{ijk},\bar{y}_{ijk})$ 
by replacing $\alpha$ and $y_{ijk}$ ($\bar{y}_{ijk}$) by $\tilde \alpha$,
$\tilde y_{ijk}$ ($\tilde{\bar{y}}_{ijk}$), i.e.,
\begin{equation}
\label{eq:spurion-RG}
\mu \frac{d \tilde \alpha}{d \mu} = \beta_\alpha(\tilde \alpha, \tilde
y_{ijk}, \tilde{\bar{y}}_{ijk}), \qquad 
\mu \frac{d \tilde y_{ijk}}{d \mu} = \beta_{y_{ijk}}(\tilde \alpha, \tilde
y_{ijk}, \tilde{\bar{y}}_{ijk}).
\end{equation}
That implies that the beta-function of the gaugino mass $M_{1/2}$ is 
obtained as 
\begin{equation}
\mu \frac{dM_{1/2}}{d\mu} = \left( M_{1/2}\alpha 
\frac{\partial}{\partial \alpha}  - a_{ijk}  
\frac{\partial}{\partial y_{ijk}}  \right) \left(  
\frac{\beta_\alpha}{\alpha}\right)
\equiv
D_1 \left(  
\frac{\beta_\alpha}{\alpha}\right).
\end{equation}

The RG equation for the soft scalar mass $m_i$ of a
chiral superfield $\phi_i$ is also easily obtained as
\begin{equation}
\mu \frac{d m^2_i}{d \mu} =
\left. 
\gamma_i(\tilde{\alpha}, \tilde{y}_{ijk}, \tilde{\bar{y}}_{ijk})
\right|_{\theta^2 \bar{\theta}^2}.
\end{equation}
These equations are found to be consistent with
the equations for the $\theta^2 \bar{\theta}^2$
components of Eqs. (\ref{eq:spurion-RG}).
Explicitly, the RG equations are written down as
\begin{eqnarray}
\mu \frac{d m^2_i}{d \mu} &=& D_2 \gamma_i \ ,\\
D_2 &=& D_1 \bar{D}_1 
+ (|M_{1/2}|^2 + \Delta_g)\alpha \frac{\partial}{\partial \alpha} 
\nonumber \\
& &
+ \frac{1}{2}(m^2_i + m^2_j + m^2_k)
\left(
y_{ijk }\frac{\partial}{\partial y_{ijk}}
+
\bar{y}_{ijk }\frac{\partial}{\partial \bar{y}_{ijk}}
\right) .
\end{eqnarray}

It is found that these RG equations lead to
very interesting properties of the soft SUSY 
breaking parameters at the vicinity of an 
IR attractive fixed point 
\cite{Karch:1998qa,Kobayashi:2001kz,Terao:2001jw}.
Deviations of the gauge coupling  and
the Yukawa coupling  from their fixed
point values, $\delta \alpha = \alpha - \alpha^*$ and 
$\delta y_{ijk} = y_{ijk} - y^*_{ijk}$, decrease exponentially.
Then the spurion method tells that both of
\begin{eqnarray}
\delta \tilde{\alpha} &=& \alpha^* M_{1/2}\theta^2
- F(\alpha^*)
\sum_i T_i m^2_i \theta^2 \bar{\theta}^2, \\
\delta \tilde{y}_{ijk} &=&
-a_{ijk} \theta^2
+\frac{1}{2}(m^2_i + m^2_j + m^2_k)y_{ijk}^*  \theta^2 \bar{\theta}^2 ,
\end{eqnarray}
also decrease exponentially towards the IR direction.
Therefore, the gaugino mass $M_{1/2}$ and the A-term $a_{ijk}$ 
are found to be suppressed\footnote{That implies that the ratio 
$a_{ijk}/y_{ijk}$ is also suppressed exponentially, 
because the Yukawa coupling $y_{ijk}$ has a fixed point.} 
and the soft scalar masses satisfy the
IR sum rules given by $\sum_i T_i m^2_i=0$ and 
$m^2_i + m^2_j + m^2_k=0$.

It is easy to see the above mentioned behavior in 
the case with a single gauge coupling only.
We consider the perturbation around the fixed point as 
$\alpha=\alpha^* + \delta \alpha$, where $\delta \alpha \ll 1$.
The beta-function of $\delta \alpha$ around the fixed point
is written as 
\begin{equation}\label{eq:RG-da}
\mu \frac{d \delta \alpha}{d \mu} = 
\left( \frac{\partial \beta_{\alpha}}
{\partial {\alpha}}  
\right)_{\alpha=\alpha^*} \delta \alpha
\equiv
\Gamma \delta \alpha.
\end{equation}
Because this fixed point is the IR attractive, that leads to 
$\Gamma > 0$.
Then, the spurion method leads immediately 
to the RG flow of the gaugino mass, that is,   
the gaugino mass is renormalized as
\begin{equation}\label{eq:M-RG-CFT}
M_{1/2}(\mu) = M_{1/2}(\mu_0) 
\left( \frac{\mu}{\mu_0}\right)^{\Gamma}.
\end{equation}
Thus the gaugino mass $M_{1/2}$ is found to be
exponentially suppressed around the IR fixed point.
Similarly, we can show that the sum $\sum_i T_i m^2_i$ 
is exponentially suppressed in this theory.
Furthermore, it is straightforward to extend this discussion 
to the theory with a gauge coupling and Yukawa couplings 
and to show that the gaugino mass $M_{1/2}$ and the A-term $a_{ijk}$ 
as well as the sums $\sum_i T_i m^2_i$ and 
$m^2_i + m^2_j + m^2_k$ are exponentially suppressed.

For the dual gauge theory with the dual quarks $q$ and
$\bar{q}$ and the meson field $M$, the superpotential
is given by $y \bar{q} M q$. Therefore the second
sum rule is 
reduced to be $m_q^2 + m_{\bar{q}}^2 + m^2_M = 0$.
We may also understand this as follows.
For example, when we use the one-loop anomalous dimensions, 
we can show that at the fixed point the gauge coupling and 
Yukawa coupling are related as
$y^*=Cg^*$, where $C$ is a constant determined by 
group-theoretical factors \cite{Pendleton:1980as}.
At the fixed point, this relation is realized as 
the relation between superfield couplings as 
$|\tilde y|^2/(8 \pi^2) = C^2 \tilde \alpha$, 
and their $\theta^2 \bar \theta^2$-terms lead to 
\cite{Kawamura:1997cw,Kobayashi:2000di}
\begin{equation}\label{eq:sum-rule}
m^2_q+m^2_{\bar q}+m^2_M = |M_{1/2}|^2.
\end{equation}
Since the gaugino mass $M_{1/2}$ is exponentially 
damping toward the conformal fixed point, the sum 
$m^2_q+m^2_{\bar q}+m^2_M$ is also exponentially damping 
as mentioned above.

\subsection{Cascading theory}

Applying the above spurion method to the cascading theory, 
we investigate the RG flow of soft SUSY breaking terms 
through the duality cascade.
We consider the $SU(kN) \times SU((k-1)N)$ gauge theory 
with two pairs of chiral matter fields $Q_r$ and $\bar Q_s$ and their 
superpotential (\ref{eq:W-4}).
The beta-functions of their gaugino masses, $M_{1/2}^{(k)}$ and 
$M_{1/2}^{(k-1)}$, are written as 
\begin{eqnarray}
\mu \frac{d M_{1/2}^{(k)}}{d \mu} &=&  
-N (k+2+2(k-1)\gamma_Q) H'(\alpha_k) \alpha_k M_{1/2}^{(k)}
\nonumber \\
& &
- 2(k-1)N H(\alpha_k)
\frac{\partial \gamma_Q}{\partial \alpha_k} 
\alpha_k M_{1/2}^{(k)}  
\nonumber \\
& &
- 2(k-1)N H(\alpha_k)
\frac{\partial \gamma_Q}{\partial \alpha_{k-1}} 
\alpha_{k-1} M_{1/2}^{(k-1)}, 
\label{M(k)beta}
\\
\mu \frac{d M_{1/2}^{(k-1)}}{d \mu} &=&  
- N (k-3+2k\gamma_Q) H'(\alpha_{k-1}) \alpha_{k-1} M_{1/2}^{(k-1)}
\nonumber \\
& &
- 2kN H(\alpha_{k-1})
\frac{\partial \gamma_Q}{\partial \alpha_{k-1}} 
\alpha_{k-1} M_{1/2}^{(k-1)}  
\nonumber \\
& &
- 2kN H(\alpha_{k-1})
\frac{\partial \gamma_Q}{\partial \alpha_k} 
\alpha_{k} M_{1/2}^{(k)}, 
\label{M(k-1)beta}
\end{eqnarray}
where $H(\alpha)=F(\alpha)/\alpha \approx \alpha$ and 
$H'(\alpha) = d H/d\alpha$.

As in Section \ref{sec:rigid-SUSY}, 
we start the RG flow at the energy scale $\Lambda$ from 
the vicinity of the fixed point A, i.e. 
$(g_k,g_{k-1},\eta) =(0,g_{k-1}^*,0)$.
Around the fixed point A, we have $k-3+2k\gamma_Q \approx 0$.
As long as $g_{k-1}$ is large and stable, the second
term in (\ref{M(k-1)beta}) reduces 
the gaugino mass $M_{1/2}^{(k-1)}$ exponentially
as the energy scale $\mu$ decreases.
On the other hand, we find $\beta_{M_{1/2}^{(k)}}<0$ 
because 
$k+2+2(k-1)\gamma_Q > 0$ and $H(\alpha_k) \approx \alpha_k \approx 0$.
Thus, the gaugino mass $M_{1/2}^{(k)}$ increases as 
the energy scale $\mu$ decreases.
However, such increase of $M_{1/2}^{(k)}$  is not drastic 
during the weak coupling region of $\alpha_k$.

Next, the theory moves from the vicinity of the fixed point 
$(g_k,g_{k-1},\eta) =(0,g_{k-1}^*,0)$ towards another fixed point, 
$(g_k,g_{k-1},\eta) =(g_{k}^*,0,0)$, where 
$k+2+2(k-1)\gamma_Q \approx 0$.
Around the latter fixed point, we find 
$\beta_{M_{1/2}^{(k-1)}}>0$ because $k-3+2k\gamma_Q < 0$, 
$H(\alpha_{k -1}) \approx 0$ and $M_{1/2}^{(k)}$ becomes irrelevant as below.
That is, the gaugino mass  $M_{1/2}^{(k-1)}$ decreases
perturbatively as the energy scale $\mu$ decreases.

On the other hand, the gaugino mass $M_{1/2}^{(k)}$ would be
suppressed exponentially in turn due to the second term in
(\ref{M(k)beta}), as 
going towards the IR fixed point.
However we note that the third term cannot be neglected,
when $\alpha_k M_{1/2}^{(k)}$ is reduced to be
comparable with $\alpha_{k-1} M_{1/2}^{(k-1)}$.
Then the gaugino mass $M_{1/2}^{(k)}$ does not follow a simple
exponential suppression. Rather it converges to a certain
value determined by $\alpha_{k-1}$ and $M_{1/2}^{(k-1)}$
obtained at the renormalized scale.

If we admit using the one-loop anomalous dimension,
then the RG behavior discussed above could be
explicitly examined.
Here we shall look into the theory on the renormalized
trajectory given in Fig.~1.
In Fig.~2, the gaugino masses $M_{1/2}^{(k-1)}(\mu)$
and $M_{1/2}^{(k)}(\mu)$ of the theory with $k=5$
are plotted with respect to
the scale parameter $\ln (\mu/\mu_0)$.
At the scale $\mu_0$, the gauge couplings are chosen
as $(\alpha_k, \alpha_{k-1}) = (0.0128, 0.04)$, which
is a point on the renormalized trajectory rather close to 
the fixed point A in Fig.~1.
The initial values at $\mu = \mu_0$ are taken to be
1.0 for both gaugino masses.

\begin{figure}[hbt]
\begin{center}
\includegraphics[width=0.65\textwidth]{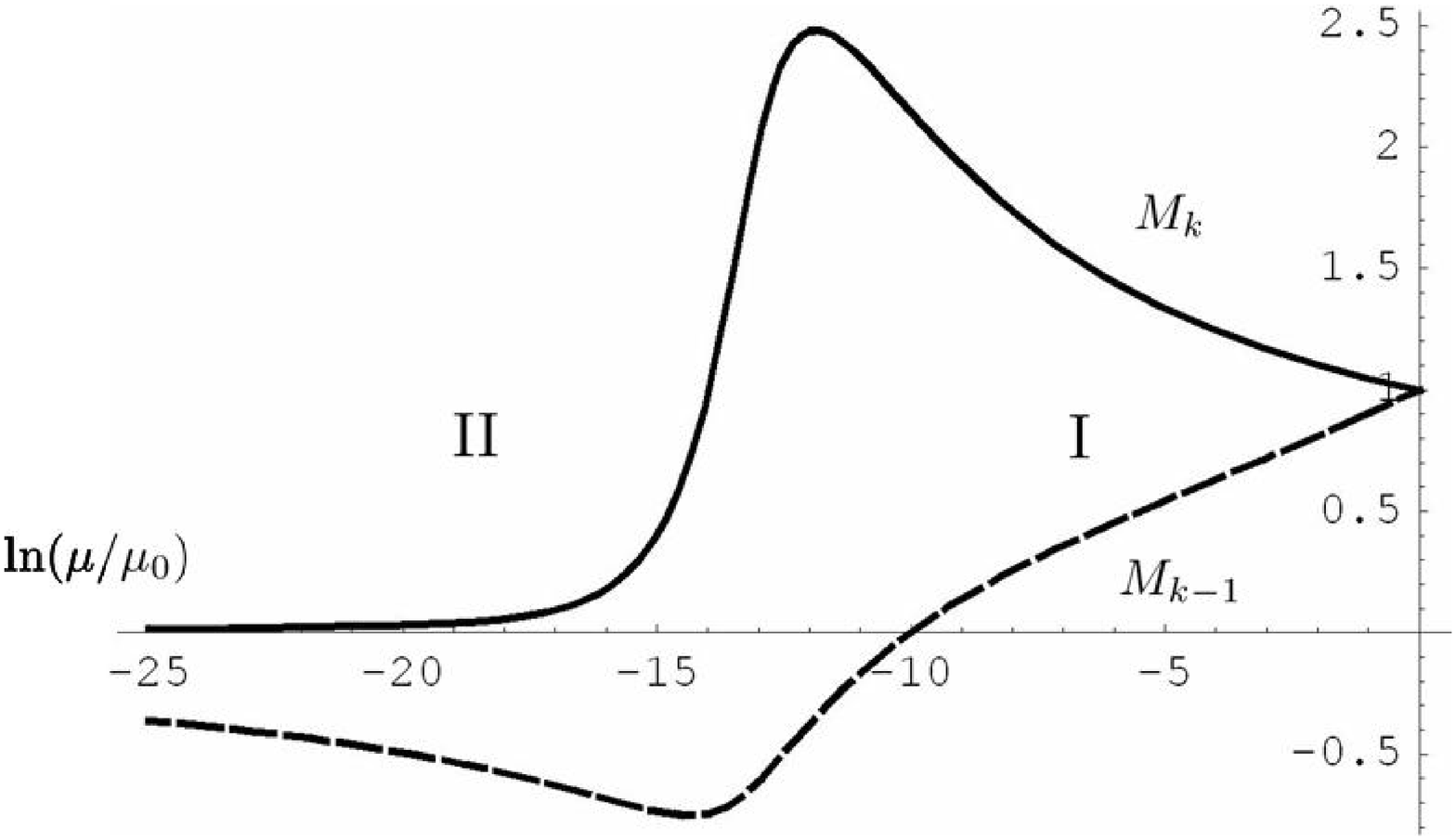}
\caption{RG running of the gaugino masses 
$M_{1/2}^{(k-1)}(\mu)$ and $M_{1/2}^{(k)}(\mu)$}
with respect to $\ln (\mu/\mu_0)$.
The gauge couplings are given at $\mu = \mu_0$
as $(\alpha_k, \alpha_{k-1})$$=$$(0.0128, 0.04)$
and  run along the renormalized trajectory.
\end{center}
\end{figure}

It is seen that $M_{1/2}^{(k-1)}$ is reduced as discussed,
but turns to be negative due to the third term in (\ref{M(k-1)beta}),
since $M_{1/2}^{(k)}$ glows slightly first.
In the region II, the gaugino mass $M_{1/2}^{(k)}$ turns out to be
suppressed strongly, while $M_{1/2}^{(k-1)}$
changes perturbatively.
In Fig.~3, the log-plot of the gaugino mass $M_{1/2}^{(k)}$
is shown by the bold line.
It is also seen that the suppression behavior deviates from
the exponential one in the end and
$M_{1/2}^{(k)}$ converges to a line.
Indeed, 
the convergence point of $\alpha_k M_{1/2}^{(k)}$ 
could be estimated as 
\begin{equation}\label{eq:converge-gaugino}
\alpha_k M_{1/2}^{(k)} \sim - 
\alpha_{k -1} M_{1/2}^{(k-1)}.
\end{equation}

\begin{figure}[hbt]
\begin{center}
\includegraphics[width=0.65\textwidth]{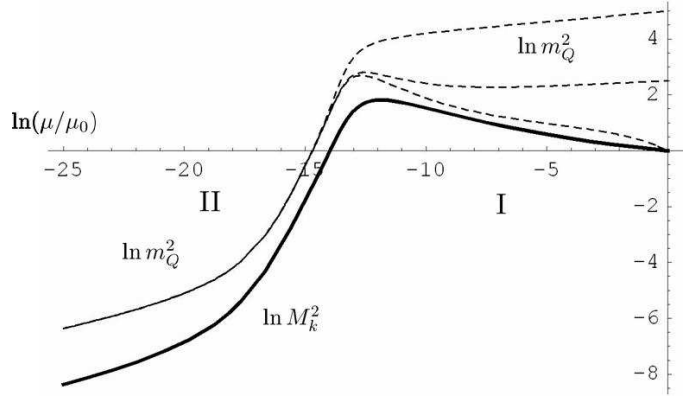}
\caption{RG running behaviors of the scalar mass 
$\ln m^2_Q$ and the gaugino mass 
$2 \ln M_{1/2}^{(k)}$ are shown by dotted lines and
the bold line, respectively.
}

\end{center}
\end{figure}

Similarly, we examine the RG running of 
the soft mass squared $m^2_Q$.
At the vicinity of the fixed points, $m^2_Q$ is 
also expected to be exponentially suppressed
as discussed in section \ref{sec:3-1}.
However existence of two gauge couplings
makes the situation more complicated.
The RG equation for $m^2_Q$ is given as
\begin{equation}
\mu \frac{d m^2_Q}{d \mu} = 
\left.
\gamma_Q(\tilde{\alpha}_k, \tilde{\alpha}_{k-1})
\right|_{\theta^2 \bar{\theta}^2}.
\end{equation}
Here, let us use the one-loop anomalous dimension
given by (9).
Then the RG equation is reduced to be
\begin{equation}
\mu \frac{d m^2_Q}{d \mu} = 
-k \alpha_k (2 |M_{1/2}^{(k)}|^2 + \Delta_k)
-(k-1) \alpha_{k-1} (2 |M_{1/2}^{(k-1)}|^2 + \Delta_{k-1}),
\label{RGeq:mQ2}
\end{equation}
where 
\begin{eqnarray}
\Delta_k &=&
H(\alpha_k)\left[
3k|M_{1/2}^{(k)}|^2 - 2(k-1)m^2_Q
\right], \\
\Delta_{k-1} &=&
H(\alpha_{k-1})
\left[
3(k-1)|M_{1/2}^{(k-1)}|^2 - 2k m^2_Q
\right].
\end{eqnarray}

In Fig.~3, the RG evolution of $m^2_Q$ of the same theory
on the renormalized trajectory is shown by
dotted lines. The initial values are taken as
$\ln m^2_Q = 0, 2.5, 5.0$ just for the illustration.
In the region I, we may neglect subleading terms of $\alpha_k$ and also
$M_{k-1}$, since it is suppressed. 
Then, Eq. (\ref{RGeq:mQ2}) is approximated to be
\begin{equation}
\mu \frac{d m^2_Q}{d \mu} \simeq
2k(k-1) (\alpha_{k-1}^*)^2 m^2_Q
- 2k\alpha_k |M_{1/2}^{(k)}|^2.
\end{equation}
This equation tells us that $m^2_Q$ is not just suppressed
but converges as
\begin{equation}
m^2_Q \to \frac{1}{(k-1) (\alpha_{k-1}^*)^2} \alpha_k |M_{1/2}^{(k)}|^2,
\end{equation}
since running of $\alpha_k |M_{1/2}^{(k)}|^2$ is rather slow.
In the case of $k=5$, the fixed point coupling
$g_{k-1}^*$ is not large and the convergence is not
so strong.
In the region II, running of $M_{1/2}^{(k)}$ changes to
exponential suppression. However, similarly to the behavior
in the region I, it converges in the IR limit as
\begin{equation}\label{eq:converge-scalar}
m^2_Q \to \frac{1}{k (\alpha_{k}^*)^2} 
\alpha_{k-1} |M_{1/2}^{(k-1)}|^2.
\end{equation}

We summarize the RG flow of the gaugino masses and soft scalar masses 
as the theory moves from the fixed point 
$(g_k,g_{k-1},\eta) =(0,g_{k-1}^*,0)$ toward the fixed point
$(g_k,g_{k-1},\eta) =(g_{k}^*,0,0)$.
At the first stage, i.e. the perturbative regime of $\alpha_k$, 
the gaugino mass $M_{1/2}^{(k-1)}$ is suppressed, while
$M_{1/2}^{(k)}$ and the soft scalar mass squared 
$m^2_Q$ increase perturbatively.
In entering the conformal regime of $\alpha_k$, 
both $M_{1/2}^{(k)}$ and $m^2_Q$ begin exponential damping,
while $M_{1/2}^{(k-1)}$ runs perturbatively.
In the IR limit, the gaugino mass $M_{1/2}^{(k)}$
and the soft scalar mass squared $m^2_Q$ are found to converge
to certain values determined by $\alpha_{k-1}$ and
$M_{1/2}^{(k-1)}$. 
Hence, these parameters evolute to be
of the same order and are fixed in the
IR limit irrespectively of their initial values.

In addition to the gaugino masses $M_{1/2}^{(k)}$, $M_{1/2}^{(k-1)}$ 
and scalar mass $m_Q$, the SUSY breaking terms corresponding to 
the superpotential (\ref{eq:W-4}) may be important, that is,
\begin{equation}\label{eq:Aterm-4}
W=h(1 - A_h \theta^2)~{\rm tr} \det_{r,s} (Q_r \bar Q_s).
\end{equation}
The RG flow behavior of the coupling $\mu hA_h$ is 
drastic following the anomalous dimensions of $Q_r$ and 
$\bar Q_s$.
Both RG flows of $\mu h$ and $\mu hA_h$ are almost the same.
That implies that their ratio $A_h$ does not change
drastically\footnote{Note that the RG flow of $\eta$ has no 
fixed point with a finite value of $\eta$.
In our case, the RG flow of $A_h$ will be ruled by
gauge couplings and gaugino masses which can be finite values. In the region II, 
$A_h$ will be affected by mainly $\alpha_{k-1}^nM_{1/2}^{(k-1)}$ in the dimensionful parameters.}.

The theory around the fixed point 
$(g_k,g_{k-1},\eta) =(g_{k}^*,0,0)$ would be well-described by 
its dual theory with the gauge coupling $g_{k-2}$ and 
the Yukawa coupling $y$.
The dual theory has the gaugino mass $M_{1/2}^{(k-2)}$, 
soft scalar masses of $q_r$, $\bar q_s$ and 
$M_{rs}$ as $m_q$ and $m_M$, the A-term $a$  and the B-term $b$.
The latter two terms are associated with the superpotential 
(\ref{eq:W-dual}) and are written as 
\begin{equation}
W=y (1 - A_y\theta^2){\rm tr}~\bar q_r M_{rs} q_s 
+ m(1 - B\theta^2){\rm tr} \det_{r,s} M_{rs}.
\end{equation}
Here, we denote $a = yA_y$ and $b=mB$.
The exact matching relations of soft terms between dual theories 
are not clear, but we assume that 
$M_{1/2}^{(k)}(\Lambda_k) \sim M_{1/2}^{(k-2)}(\Lambda_k)$ and 
all of soft scalar masses are of the same order at $\Lambda_k$.
Furthermore, we assume that all of $A_h$, $A_y$ and $B$ are 
of the same order at $\Lambda_k$.

When the gauge coupling $g_{k-2}$ approaches 
toward its non-trivial fixed point, the gaugino mass 
$M_{1/2}^{(k-2)}$ and soft scalar mass squared $m^2_q$ are 
also exponentially suppressed.
This behavior is similar to that of $M_{1/2}^{(k)}$
and $m^2_Q$ discussed previously.
Moreover, in the dual theory the Yukawa coupling $y$ 
approaches to the fixed point $y^*$.
In this case, a small deviation $\delta y = y - y^*$ 
is exponentially damping as (\ref{eq:RG-da}).
The spurion method leads that the A-term coupling $A_y$ is also
suppressed exponentially.
On the other hand, the RG behavior of $B$ is rather similar to 
one of $A_h$.
It is found that both RG flow behaviors of 
the mass $m/\mu$ and $b/\mu$ are almost the same and 
they are determined by large anomalous dimensions of 
$M_{rs}$.
However, their ratio $B$ does not change drastically\footnote{Note
 that the RG flow of $m/\mu$ has no fixed point with its fine value.
In this IR region, $B$ will be affected by mainly $\alpha_{k-1}^nM_{1/2}^{(k-1)}$ 
in the dimensionful paramters.}.

In the dual theory, not only $m^2_q$ but also
the sum of soft scalar masses squared, 
$m^2_q+m^2_{\bar q}+m^2_M$, is also suppressed
in the conformal region.
That implies that each of $m^2_q$ and $m^2_M$
is suppressed when $m^2_q =m^2_{\bar q}$, which is the 
relation we are assuming. 
However, we cannot neglect the effects through
$SU(N(k-1))$ gauge interaction such as the discussions of 
convergence points, (\ref{eq:converge-gaugino}) and 
(\ref{eq:converge-scalar}), 
in the original $SU(Nk) \times SU(N(k-1))$ theory.

The gaugino mass $M_{1/2}^{(k-2)}$, the A-parameter $A_y$ 
and the scalar masses squared $m^2_q$ and $m^2_M$ in the dual theory
are not just suppressed out, rather converge to certain
values given by $\alpha_{k-1}$ and 
$M_{1/2}^{(k-1)}$ in the IR limit again.
It is straightforward to solve the RG equations,
if we admit using the one-loop anomalous dimensions
of $q$ and $M$ just as performed above.
However, we shall avoid to present a similar 
analysis here.
It may be explicitly seen that both $m^2_q$ and 
$m^2_M$ as well as $M_{1/2}^{(k-2)}$ and $|A_y|^2$
converge the values of the same order
given by $\alpha_{k-1}|M_{1/2}^{(k-1)}|^2$.
The meson field $M$ belongs to the adjoint
representation of $SU(N(k-1))$ group and
suffers from the effects through
$SU(N(k-1))$ gauge interaction more.
Therefore, $m^2_M$ is found to be positive
and larger than $m^2_q$ in the IR
\footnote{Soft masses for singlet mesons $M^0_{rs}$
may be driven to be negative because of the Yukawa couplings.}.


When the supersymmetric mass $m$ of the chiral fields $M_{rs}$ 
becomes large, we integrate out these fields.
Then, the low energy theory becomes the 
$SU((k-1)N) \times SU((k-2)N)$ gauge theory with two 
pairs of bifundamental and anti-bifundamental fields and 
the quartic superpotential, $W = \tilde h {\rm tr} \det q_r \bar q_s$.
The theory has soft SUSY breaking terms, i.e. the gaugino masses, 
$M_{1/2}^{(k-1)}$ and $M_{1/2}^{(k-2)}$, and soft scalar mass 
$m_q$.
In addition, we have the SUSY breaking term corresponding to 
the superpotential $W = \tilde h {\rm tr} \det q_r \bar q_s$, that is, 
\begin{equation}
W = \tilde h (1 - \theta^2 A_{\tilde h}) {\rm tr} \det q_r \bar q_s .
\end{equation}
The size of $A_{\tilde h}$ may be of the order of $B$ or 
$A_y$ at the decoupling scale of $M_{rs}$.
If these SUSY breaking terms are smaller than other mass scales 
such as the energy scale $\mu$ and the supersymmetric mass $m$, 
the above cascade continues as rigid SUSY theory in 
Section \ref{sec:rigid-SUSY}.
Through the cascade, 
the gaugino masses and soft scalar masses are damping 
except the perturbative regime, where the theory moves from 
the fixed point $(g_k,g_{k-1},\eta) =(0,g_{k-1}^*,0)$ toward the fixed point
$(g_k,g_{k-1},\eta) =(g_{k}^*,0,0)$.
When we integrate out $M_{rs}$, which are charged under 
the $SU((k-1)N)$ gauge group, threshold corrections would appear.
For example, the gaugino mass $M_{1/2}^{(k-1)}$ would receive 
such threshold corrections $\Delta M_{1/2}^{(k-1)}$, 
which would be estimated by $\Delta M_{1/2}^{(k-1)}={\cal O}(\alpha_{k-1}B)$.
That would be small, because $\alpha_{k-1}$ is small.
At any rate, if the cascade continues, the total gaugino mass 
$M_{1/2}^{(k-1)}$ would be suppressed at the next stage 
such as the gaugino mass $M_{1/2}^{(k)}$ is suppressed at the stage 
discussed above.

As discussed above, the cascade would continue 
unless SUSY breaking terms are comparable with 
other mass scales such as the energy scale $\mu$ 
and the supersymmetric mass $m$.
Gaugino masses and SUSY breaking scalar masses would be suppressed 
through the cascade except the regime I, where 
the gaugino mass $M_{1/2}^{(k)}$ would increase.
On the other hand, the SUSY breaking parameters, $B$ and $A_h$, 
would not be suppressed like the others.
Note that the B-term corresponds to the off-diagonal entries 
of mass squared matrix of the fields $M_{rs}$, 
that is, eigenvalues of mass squared would be written by 
$|m|^2+m^2_M\pm |mB|$.
A large value of $|B|$ would induce a tachyonic mode.
Then, the scalar component of superfields $M_{rs}$ 
may develop their VEVs and the 
gauge symmetry $SU((k-1)N)$ may be broken.
Also, through this symmetry breaking, the matter fields 
$q_r$ and $\bar q_s$ may gain mass terms due to the Yukawa coupling 
with $M_{rs}$.
Then, the duality cascade would be terminated when 
mass parameters, $|m|^2$, $|mB|$ and $m_M^2$, are
comparable \footnote{Similarly, the singlet meson fields $M_{rs}^0$ 
may develop their VEVs depending on values of their various mass terms.
Their VEVs induce mass terms of dual quarks.
If such masses are large enough, the dual quarks would decouple and 
the flavor number would reduce to be outside of the conformal window.
Then, the cascade could end.
In addition, scalar components of $q_r$ and $\bar q_s$ may 
develop their VEVs depending on values the A-terms and 
their soft scalar masses as well as other parameters in the 
scalar potential.
Their VEVs break gauge symmetry and the cascade would end. 
}.
This type of ending of the duality cascade could happen 
only in the softly broken supersymmetric theories and 
such symmetry breaking would be important.
Thus, we will study such breaking more explicitly 
in the next section.
Similar symmetry breaking would be realized 
not only in the ``magnetic dual theory'', but also in 
the original ``electric theory'' with the 
quartic A-term (\ref{eq:Aterm-4}). 
If the quartic A-term is comparable with SUSY breaking scalar masses 
$m_Q$, the origin of scalar potential of $Q$ would be 
unstable and similar symmetry breaking would happen.
Such gauge symmetry breaking with reducing the flavor number 
may correspond to the symmetry breaking by VEVs of $M_{rs}$ 
with inducing dual quark masses.

Whether $M_{rs}$ include tachyonic modes depends on 
values of $|m|^2+m^2_M\pm |mB|$, i.e. their initial 
conditions as well as matching conditions.
In a certain parameter region, the scalar fields 
$M_{rs}$ may include tachyonic modes and symmetry 
breaking may happen.
In the other parameter regions, the cascade would continue
like the rigid supersymmetric theory.
For example, when the magnitude of SUSY breaking terms 
is much smaller than the supersymmetric mass $m$ and 
the energy scale $\mu$, the cascade would continue 
in almost the same way as the rigid supersymmetric theory.
Then, it would end with the pure $SU(N)$ supersymmetric Yang-Mills 
theory with non-vanishing gaugino mass.


\section{Symmetry breaking and illustrative model}
\label{sec:symm-br}

\subsection{Symmetry breaking}

In the previous section, we have pointed out the possibility 
that a tachyonic mode in the meson fields $M_{rs}$ 
would appear because of soft SUSY breaking terms and 
its VEV would break the symmetry.
Here, we study this aspect more explicitly.

\subsubsection{$SU(kN) \times SU((k-1)N)$ model}

First, we study the $SU((k-2)N)\times SU((k-1)N)$ theory, which 
is dual to the $SU(N) \times SU((k-1)N)$ theory.
As discussed in the previous section, 
the dual theory includes the meson fields 
$M_{rs}$, which have the supersymmetric mass $m$, 
the SUSY breaking soft scalar masses $m_M$ and 
the B-term $mB$.
That is, their scalar potential $V$ is written by 
\begin{eqnarray}
V &=&  (|m|^2+m_M^2)\sum_{rs}|M_{rs}|^2 + 
(mB (M_{11}M_{22} - M_{12}M_{21}) + h.c.) + V_D^{(k-1)}, \nonumber \\
V_D^{(k-1)}  &=& \frac12g^2_{k-1}D^2_{(k-1)} ,
\end{eqnarray}
where $D_{(k-1)}$ denotes the D-term of the 
$SU((k-1)N)$ vector multiplet.
Here, we have assumed the $SU(2)$ invariance for the $(r,s)$ 
indices of $M_{rs}$.
The eigenvalues of mass squared matrix are given by 
\begin{equation}\label{eq:mass-eigen}
|m|^2+m_M^2 \pm |mB|.
\end{equation}
If $|m|^2 \gg |m_M^2|, |mB| $, the theory is almost supersymmetric 
and the duality cascade would continue.
(Note that $m$ is the supersymmetric mass and the others 
are masses induced by SUSY breaking.) 
However, if the masses (\ref{eq:mass-eigen}) include 
a negative eigenvalue, there appears a tachyonic mode at 
the origin of the field space $M_{rs}$.
Note that the D-flat direction corresponds to the VEV direction, 
where $M_{rs}$ are written by diagonal elements, 
that is, Cartan elements.
That implies that when a negative eigenvalue is included 
in (\ref{eq:mass-eigen}), such a direction would be 
unbounded from below in the tree-level scalar potential.
Thus, the meson fields $M_{rs}$ would develop their VEVs,
whose order would be equal to the cut-off scale 
of the $SU((k-2)N) \times SU((k-1)N)$ theory, i.e. 
$\Lambda_k$.
The VEVs of adjoint fields $M_{rs}$ break the gauge group $SU((k-1)N)$ 
to a smaller group and induce mass terms of 
$q_r$ and $\bar q_s$ through the Yukawa couplings 
$yq_rM_{rs}\bar q_s$.

\subsubsection{$\prod_i SU(N_i)$ quiver model }

\begin{figure}
\begin{center}
\includegraphics[scale=0.3]{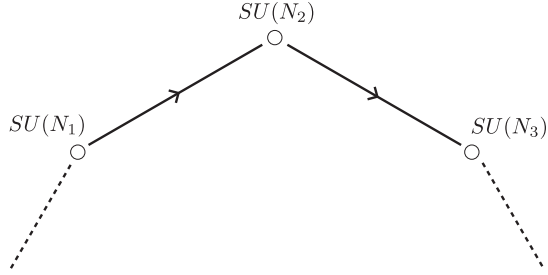}
\end{center}
\caption{$\prod_i SU(N_i)$ quiver model}
\label{fig:suni}
\end{figure}

The above analysis can be extended to the $\prod_i SU(N_i)$ 
quiver gauge theory with their bifundamental matter fields.
We consider a subsector of the quiver theory, that is, 
the $SU(N_1) \times SU(N_2) \times SU(N_3)$ gauge theory 
with bifundamental matter fields, 
$(N_1,\bar N_2,1)$ and $(1,N_2,\bar N_3)$ as shown in Fig.~\ref{fig:suni}.
The $SU(N_1)$ and $SU(N_3)$ sectors would have other types of 
bifundamental matter fields, but we neglect them
\footnote{In each non-abelian gauge group, for example, we need vector-like matter fields 
in order to cancel anomaly. However, we assume that the theory is anomaly-free 
at every stage.}.
In addition, for simplicity we consider the case with 
$N_1 = N_3$.
Here, we dualize the $SU(N_2)$ sector.
Then, there appear the dual matter fields $q$ and $\bar q$ with
the representations $(\bar N_1,\tilde N_2,1)$ 
and $(1,\bar {\tilde N_2},N_3)$, where $\tilde N_2 = N_1 - N_2$.
In addition, the meson field $M$ with the 
representation $(N_1,1,\bar N_3)$ appears 
and has Yukawa couplings among $q$ and $\bar q$.
The supersymmetric mass term of the meson field 
in the superpotential 
is not allowed, i.e. $m=mB=0$.
In this case, only the SUSY breaking soft scalar mass $m_M$ 
as well as the D-term potentials  appears 
in the scalar potential of the meson field $M$.
Thus, the scalar potential is simple.
The scalar mass squared $m_M^2$ tends to converge to a positive value 
as discussed in the previous section.
Thus, the symmetry breaking may not happen by 
the VEV of $M$ in this theory.

When $N_1 = N_3 =2$, supersymmetric mass terms of meson fields 
in the superpotential would be allowed.
Alternatively, when the model includes anti-meson fields 
$\bar M$, the supersymmetric mass term would be allowed 
in the superpotential.
In these models, the corresponding B-terms would also be allowed.
Furthermore, in the latter model, there are D-flat directions, 
i.e. $M= \pm \bar M$.
In this case, the scalar potential would be written by 
\begin{eqnarray}
V &=&  (|m|^2+m_M^2)|M|^2 + (|m|^2+m_{\bar M}^2)|\bar M|^2 + 
(mB M\bar M + h.c.) \nonumber \\ 
& & + V_D^{(N_1)}+ V_D^{(N_3)}, 
\end{eqnarray}
where $V_D^{(N_1)}$ and $V_D^{(N_3)}$  are D-term scalar potentials 
for the $SU(N_1)$ and $SU(N_3)$ vector multiplets.
This potential at the tree level is unbounded from below along the 
D-flat direction $M= \pm \bar M$ if 
\begin{equation}
2 |m|^2 + m_M^2 + m_{\bar M}^2 < 2|mB|.
\end{equation}
In addition, the meson fields include a tachyonic mode 
when 
\begin{equation}
(|m|^2+m_M^2) (|m|^2+m_{\bar M}^2) < |mB|^2,
\end{equation}
or 
\begin{equation}
(|m|^2+m_M^2) (|m|^2+m_{\bar M}^2) > |mB|^2 
{\rm~~and~~} 2 |m|^2 + m_M^2 + m_{\bar M}^2 < 0.
\end{equation}
Thus, various phenomena could happen depending on 
values of mass parameters, $m$, $m_M$, $m_{\bar M}$ and $mB$, 
that is, the unbounded-from-below direction, 
the symmetry breaking without the unbounded-from-below direction 
or no symmetry breaking.
Indeed, this situation is quite similar to what 
happens in the Higgs scalar potential of 
the minimal supersymmetric standard model (MSSM).

\subsection{Illustrating model}

Here we give a simple example of theories, whose 
field contents are  similar to 
the MSSM or its extensions and where symmetry breaking 
would happen.


\begin{figure}
\begin{tabular}{ccc}
\begin{minipage}{0.4\hsize}
\includegraphics[width=\hsize]{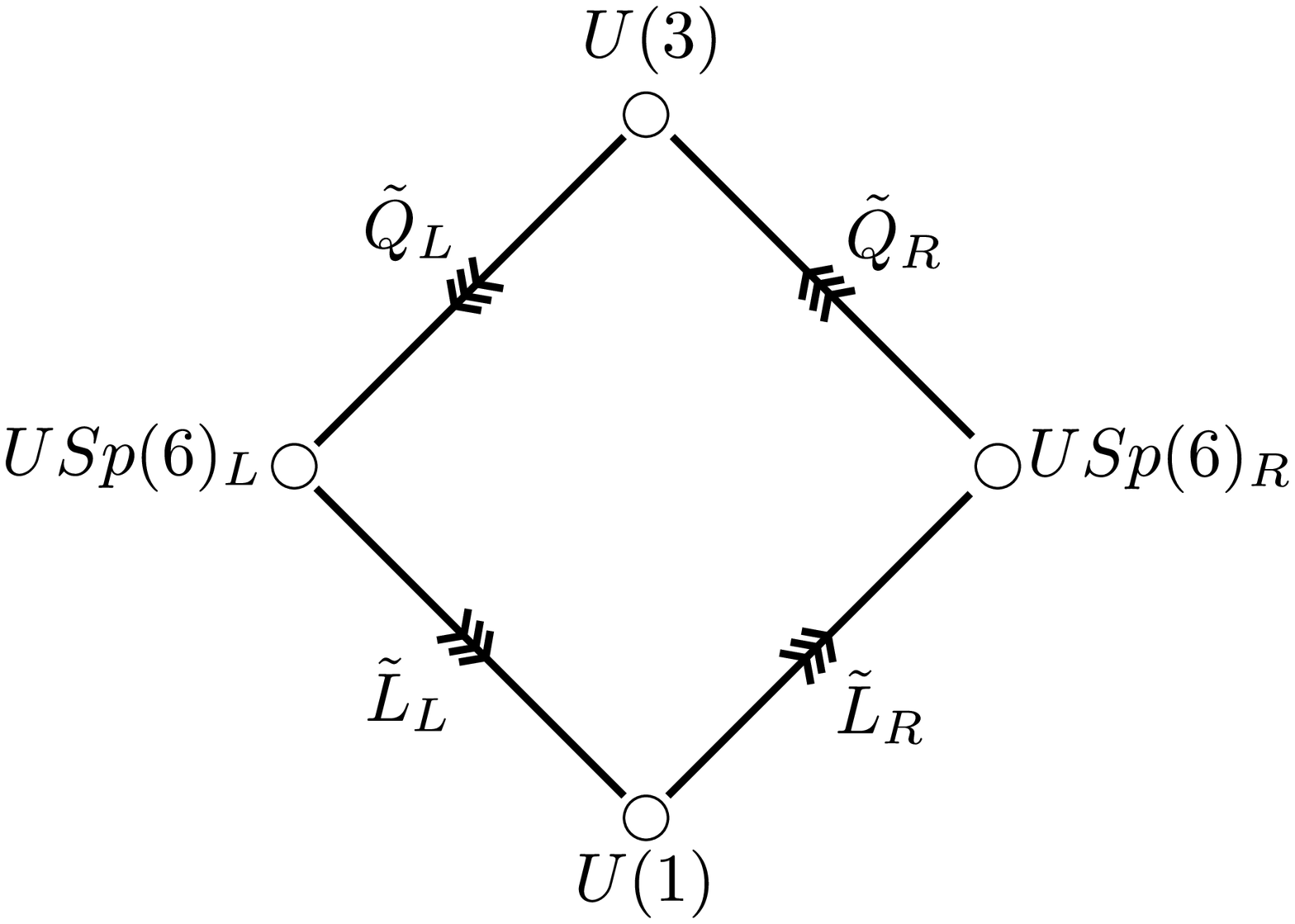}
\end{minipage}
&\begin{minipage}{0.05\hsize}$\rightarrow$\end{minipage}&
\begin{minipage}{0.4\hsize}
\includegraphics[width=\hsize]{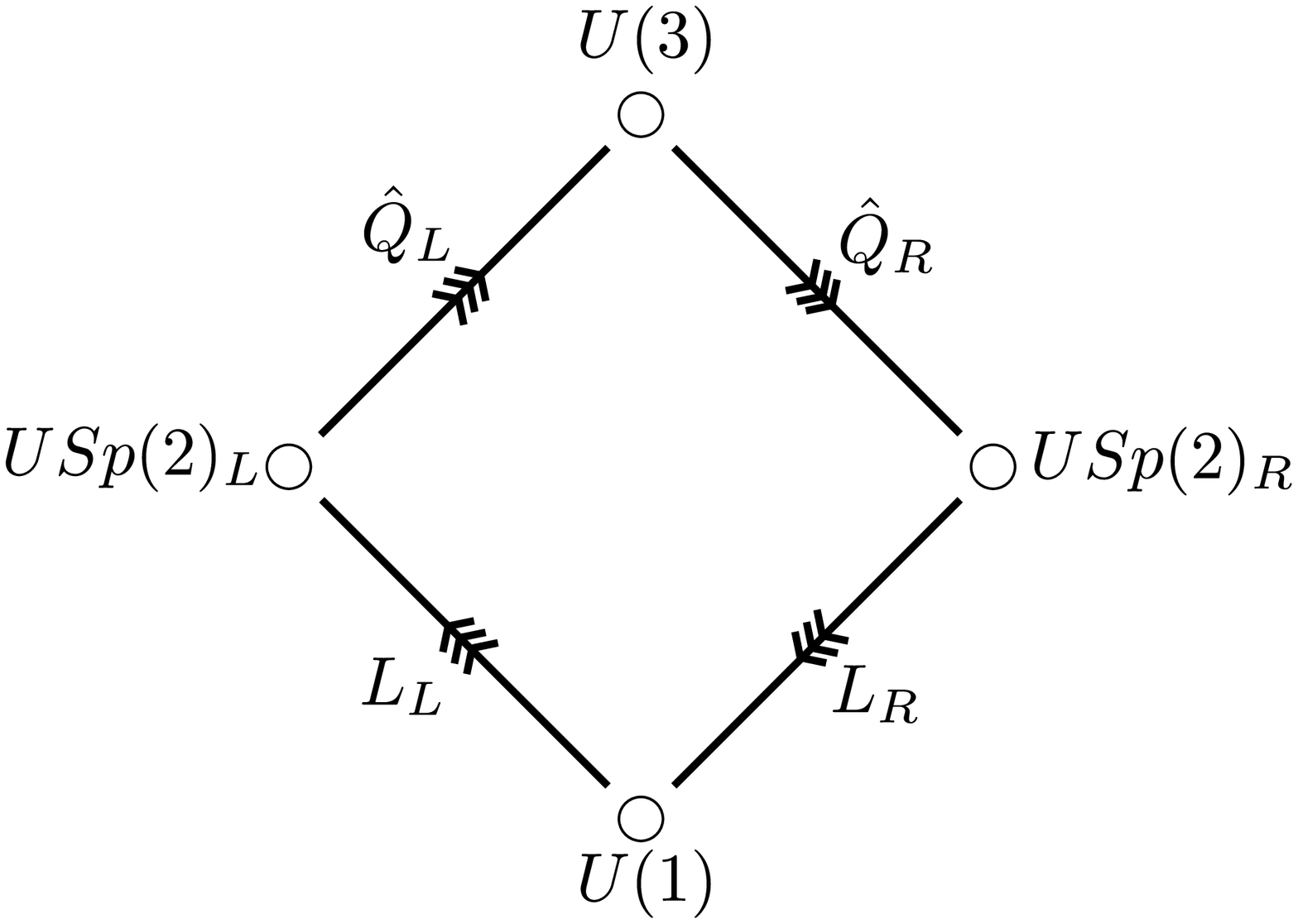}
\end{minipage}\\
&&\\
&\begin{minipage}{0.05\hsize}$\rightarrow$\end{minipage}&
\begin{minipage}{0.4\hsize}
\includegraphics[width=\hsize]{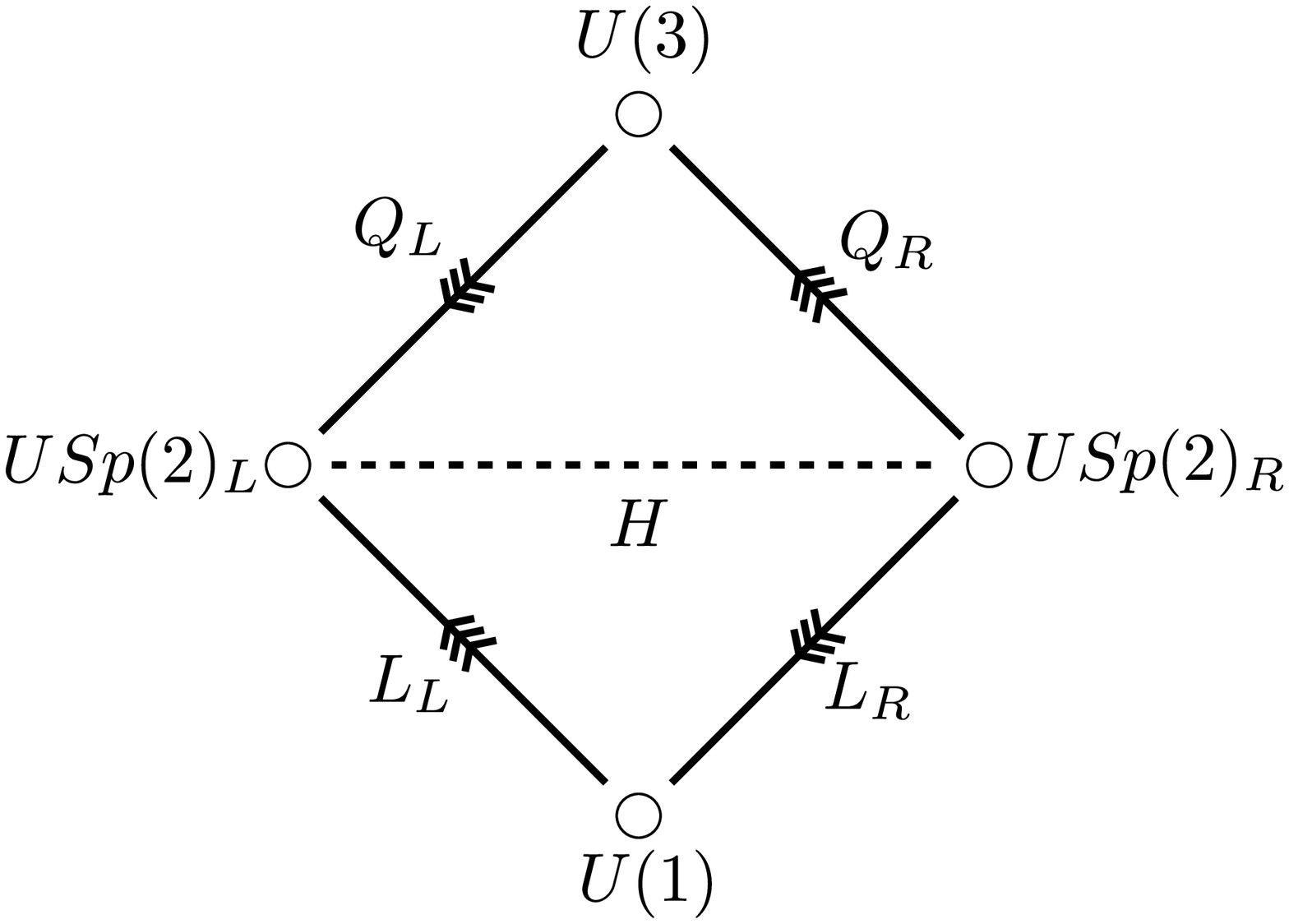}
\end{minipage}
\end{tabular}
\caption{Quiver diagrams of the illustrating model}
\label{fig:u3}
\end{figure}

We consider the gauge group 
$U(3)\times USp(6)_L \times USp(6)_R \times U(1)$ and 
three families of bifundamental fields, 
$\tilde Q_L:(3,6,1,0)$, $\tilde Q_R:(\bar 3,1,6,0)$, $\tilde L_L:(1,6,1,-1)$ 
and $\tilde L_R:(1,1,6,1)$ and the superpotential 
\begin{equation}
W = h \tilde Q_L \tilde Q_R \tilde L_L \tilde L_R.
\end{equation}

We expect that first the gauge couplings of $USp(6)_L \times USp(6)_R$
would approach to their non-trivial fixed point.
Then, the $USp(6)_L \times USp(6)_R$ sector is dualized, that is, 
the gauge group is $U(3)\times USp(2)_L \times USp(2)_R \times
U(1)$ as shown in Fig.~\ref{fig:u3}.
Note that $USp(2) \simeq SU(2)$.
In addition we would have matter fields, 
$\hat Q_L:(\bar 3,2,1,0)$, $\hat Q_R:(3,1,2,0)$, $L_L:(1,2,1,1)$ 
and $L_R:(1,1,2,-1)$.
Also, we would have several ``meson fields'' 
$M:(3,1,1,-1)$ and $\bar M:(\bar 3,1,1,1)$, which have mass terms 
$mM\bar M$ and Yukawa couplings with $\hat Q_L, \hat Q_R, L_L$ 
and $L_R$, but they can be integrated out because of heavy mass terms 
$mM\bar M$.
Then, we obtain the superpotential
\begin{equation}\label{eq:model-1-W-2}
W = \hat h \hat Q_L \hat Q_R  L_L        L_R.
\end{equation}

Next, we expect that the gauge coupling of $SU(3)$ approaches to 
the conformal fixed point.
Then, the $U(3)$ sector is dualized.
The gauge group is $U(3)\times USp(2)_L \times USp(2)_R \times
U(1)$ and we would have matter fields, 
$Q_L:(3,2,1,0)$, $Q_R:(\bar 3,1,2,0)$, $L_L:(1,2,1,1)$ 
and $L_R:(1,1,2,-1)$ as well as several ``Higgs fields'' 
$H:(1,2,2,0)$.
The superpotential is obtained as
\begin{equation}
W = y_Q Q_L Q_R H + y_L L_L L_R H + m HH.
\end{equation}
Note that the operator (\ref{eq:model-1-W-2}) 
corresponds to $y_L L_L L_R H$.
However, the gauge symmetry $U(3)\times USp(2)_L \times USp(2)_R \times
U(1)$ allows the mass terms $mHH$.
Thus, we assume that such terms would be induced and 
we have added such terms.
Then, if SUSY breaking terms induce a tachyonic mode of $H$, 
the symmetry $USp(2)_L \times USp(2)_R$ would be broken.

In this model, $USp(2)_L$ and $USp(2)_R$ symmetry 
breaking would happen at the same time.
Although the left-right asymmetry is required for 
a realistic model, it would be difficult to generate such 
left-right asymmetry in this model.
Some modification is necessary for a realistic model. 
At any rate, this model is an illustrating model 
for symmetry breaking.
Such symmetry breaking by SUSY breaking terms in 
the duality cascade may be important, e.g. 
to realize the standard model at the bottom of the 
cascade.
We would study model building towards more realistic models elsewhere.

\section{Conclusion}
\label{sec:conclusion}

We have studied the RG flow of softly broken 
supersymmetric theories showing the duality cascade.
Gaugino masses and scalar masses are suppressed 
in most regime of the RG flow although they increase 
in a certain perturbative regime.
After exponential damping, the gaugino mass 
$M^{(k)}_{1/2}$ corresponding to the strongly 
coupled sector converges to a certain value, 
which is determined by the gauge coupling $\alpha_{k-1}$ 
and the gaugino mass $M_{1/2}^{(k-1)}$ in 
the weakly coupled sector.
The scalar mass would also converge to the same order value.
At the next stage of the cascade, 
the strongly and weakly coupled sectors are 
interchanged with each other and 
the gaugino mass $M_{1/2}^{(k-1)}$ 
would be suppressed.
Thus, through the sequential cascade, 
the magnitude of gaugino masses and scalar masses would be 
suppressed.

The B-term may be important.
In a certain parameter region, the B-term would 
induce tachyonic modes of $M_{rs}$ and symmetry 
breaking would happen.
Such an aspect would be important to realistic model building.

The RG flow of SUSY breaking terms in the cascading theory 
is quite non-trivial 
as the RG flow of gauge couplings.
The gravity dual of the cascade rigid supersymmetric theory has 
been studied extensively.
However, our analysis implies that the dilaton is also running as
$e^{-\phi} \sim \alpha_k^{-1} +  \alpha_{k-1}^{-1}$ \cite{Klebanov:2000hb},
but the supergravity solution of the D3-brane does not admit this running behavior
and most of the supergravity dual theories concentrate on the constant dilaton backgrounds.
In this sense, the suppression of the gaugino masses would be a quite different mechanism
from the suppression due to the warp factor as already pointed out in
\cite{DeWolfe:2002nn,Kachru:2003aw,Giddings:2005ff,Douglas:2007tu}.
The region of RG flow in our study might 
be outside of the supergravity approximation,
but it would be quite interesting to study the 
gravity dual side corresponding to the RG flow of 
SUSY breaking terms including the dilaton running.

We have considered the scenario that supersymmetry is 
broken at high energy and investigated the RG flow of 
SUSY breaking terms.
Alternatively, we could consider another scenario that 
supersymmetry would be broken at some stage of 
the cascade.
For example, supersymmetry is broken dynamically 
through the cascade and such breaking is mediated to 
the visible sector.
Such a study would also be important.
We would study such a scenario elsewhere.

\vspace{1cm}
\noindent
{\bf Acknowledgement}

H.~A.\/,  T.~H.\/, T.~K.\/,  K.~O.\/, Y.~O.\/ and H.~T. \/ 
are supported in part by the
Grant-in-Aid for Scientific Research, No.~182496, No.~194494, 
No.~20540266, No.~19740120, No.~$20\cdot 324$ 
and No.~17540245, from the Ministry of Education, Culture,
Sports, Science and Technology of Japan.
T.~K.\/ is also supported in part by 
the Grant-in-Aid for the Global COE Program 
"The Next Generation of Physics, Spun from Universality 
and Emergence" from the Ministry of Education, Culture,
Sports, Science and Technology of Japan.

\end{document}